# Spontaneous Transition from Conformal to Two-Dimensional Growth in Ge/GeSn Core/Shell Nanowires

S. Assali,[1, 2, ||, *] S. Koelling,[1, ||] M. Andelic,[3] L. Luo,[1] G. A. Botton,[3,4] and O. Moutanabbir[1]

[1] Department of Engineering Physics, École Polytechnique de Montréal, C. P. 6079, Succ. Centre-Ville, Montréal, Québec H3C 3A7, Canada

[2] University Grenoble Alpes, CEA, Grenoble INP, IRIG, PHELIQS, 38000 Grenoble, France

[3] Department of Materials Science and Engineering, McMaster University, Hamilton, ON, Canada

[4] Diamond Light Source, United Kingdom

GeSn semiconductors are group-IV isovalent alloys that offer remarkable tunability of optoelectronic properties across the entire infrared spectrum, while remaining fully compatible with silicon processing standards. These attributes make GeSn a promising platform for scalable sensing, imaging, and communication technologies. Yet, the influence of dimensionality on GeSn crystal growth remains poorly understood, limiting the development of integrated nanoscale infrared devices. Here, we reveal the spontaneous formation of hitherto unreported ultra-thin GeSn fins with sub-30 nm thickness during vapor-phase growth on Ge nanowire substrates. A transition from the typical conformal GeSn shell to distinct fin-like structures occurs along the nanowire growth axis and is accompanied by ordered twin defects extending longitudinally and laterally, inducing a transition from diamond to hexagonal-like crystal structure. The fins exhibit uniform Sn incorporation of approximately 16 at.% throughout their volume, indicating high compositional homogeneity. These findings uncover an anisotropic growth regime in metastable GeSn alloys, enriching the fundamental understanding of nanoscale epitaxy.

GeSn semiconductors provide additional degrees of freedom to engineer the electronic structure in silicon-compatible platforms, thereby enabling innovative approaches to implement a variety of monolithic optoelectronic, photonic, and quantum devices.[1–4] For instance, in this binary alloy, a direct bandgap semiconductor is obtained when the Sn content is increased above ~7 at.% in a strain-free material. Thanks to major advances in materials development over the course of the last decade, this characteristics has been exploited to demonstrate silicon-integrated GeSn photodetectors,[5–8] lasers,[9,10] and LEDs.[11,12] These devices cover operation in short-wave infrared (SWIR: 1.5-3 μm) and mid-wave infrared (MWIR: 3-8 μm) bands. Despite these important achievements, one of the major hurdles still facing GeSn remains the difficulty obtaining high-quality crystalline materials due to thermodynamic constraints limiting, the solid solubility of Sn in Ge to less than 1 at.%. Non-equilibrium growth methods at substrate temperatures below 400 °C are currently utilized to grow metastable, diamond cubic GeSn materials on Si using Ge as an interlayer (Ge/Si substrate). The metastable nature of these alloys brings about additional constraints in terms of phase separation leading to Sn precipitation and surface segregation that compromise the material quality.[13] This degradation may result from either deviations from optimal growth conditions[13] or post-growth processing at higher temperatures.[14,15] Moreover, compressive strain in the growing GeSn layer strongly suppresses the uniform incorporation of Sn and further limits the alloy stability.[16]

The aforementioned inherent challenges peculiar to the growth of GeSn can be partially alleviated when substrates with reduced dimensionality, such as nanowires (NWs), are used as compliant growth substrates.[17,18] NWs are typically grown using the vapor-liquid-solid (VLS) process, and the growth of GeSn NWs[19,20] and NW branches[21] from a metal catalyst was demonstrated. Moreover, extending this protocol to radial heterostructures resulted in the growth of conformal GeSn shells around a Ge core NW.[22–24] Using quasi-one-dimensional substrates allows for an efficient strain minimization during shell growth, leading to a highly uniform incorporation of Sn up to 18 at.% while preserving the crystalline quality.[23,25–28] This uniform incorporation of Sn is observed over a few hundreds of nanometers in thickness and a few micrometers in length,[28] which is very challenging to achieve in conventional planar GeSn growth on Ge-buffered Si.[16,29,30]

In this letter, we demonstrate the existence of a transition from this conformal growth to two-dimensional fins growth. Here, the growth seems to simultaneously abort along two opposite facets yielding sub-30 nm-thick GeSn fins stemming from the four remaining facets of a Ge NW core. Uniform Sn content of 16 at.% is obtained, and twinning defects, observed along both lateral and longitudinal directions of the fins, are believed to be associated with this transition. Besides expanding the fundamental understanding of epitaxy on nanoscale scale substrates, the capability to harvest the shell-fins growth transition may also lay the ground for flexible infrared optoelectronic devices.

The Ge/GeSn nanomaterials were grown from 20 nm Au colloids dispersed on a Ge (111) wafer in a chemical vapor deposition (CVD) reactor, using germane ($GeH_4$) and tin-tetrachloride ($SnCl_4$) as precursor gases. The VLS growth of the <111> Ge NWs was performed at 320 °C, while $Ge_{1-x}Sn_x$ was grown at 280 °C, using the same approach as in Ref.[28]. The Scanning Electron Microscopy (SEM) images of the NWs are shown in Fig. 1a. A large fraction of NWs displays the conventional Ge/GeSn core/shell heterostructures with limited tapering. On the same sample, however, highly bent NWs with a flat portion are also observed. These NWs exhibit a clear transition from the conformal GeSn shell growth to fin-like nanostructures. The crystalline quality of the NWs is investigated with transmission electron microscopy (TEM), while the chemical composition is evaluated using energy-dispersive X-ray spectroscopy (EDX) in cross-sectional scanning-TEM (STEM) high angle annular dark field (HAADF) imaging. In TEM of the core/shell NWs (Fig. 1b) the uniform contrast observed along the length of the NW indicates the absence of extended defects across the heterostructure. The STEM (Fig. 1c) and EDX (Fig. 1d) images demonstrate high content uniformity in the GeSn shell, without segregation of Sn on the NW sidewall.[31] Interestingly, the spontaneous transition from core/shell to fin growth was found to systematically take place at 1-2 μm from the bottom of the Ge NW, as it can be seen in a representative SEM image (Fig. 1e). The TEM image of a Ge/GeSn core/fin nanostructure (Fig. 1f) shows clear wrinkling of the surface because of its relatively small thickness, typically below 30 nm, as discussed later. Uniform contrast is again observed in the STEM (Fig. 1g) and EDX (Fig. 1h) images, thus confirming that segregation of Sn is avoided in the fin. The Sn compositional profiles, extracted along the direction perpendicular to the Ge core for both the core/shell and fin regions of the NWs, are plotted in Fig. 1k. The Ge core can easily be identified in the 200 nm-wide Ge/GeSn core/fin heterostructure, while in the core/shell NW the conformal shell growth largely overwhelms the contribution of the Ge core to the EDX signal.

Deeper insight into the spontaneous growth transition are obtained by performing cross-sectional TEM measurements. As a reference, the core/shell Ge/GeSn NW heterostructure is shown in the cross-sectional HAADF-STEM image in Fig. 2a and EDX compositional map in Fig. 2b, adapted from Ref.[27]. Sn-rich {112} wide facets and Ge-rich {110} narrow facets are observed around the 20 nm Ge core, with a typical sunburst-like morphology for the Ge/GeSn core/shell NW system.[22,23,32] To obtain an accurate estimation of the NW stoichiometry, Atom Probe Tomography (APT) measurements were performed, as displayed in Fig. 2c. We note that the APT instrument used in this work (Invizo 6000) allows the analysis with a significantly larger field-of-view exceeding 60 % of the entire NW cross-section, while providing a more accurate estimation of the chemical composition down to the ppm level.[33] The Sn content on the {112} wide facets gradually increases to 16 at.% in the outer region of the 30 nm-thick shell, while the {110} narrow facets display a more constant incorporation of 6 at.%. The compositional profile estimated along the <112> radial direction of the shell from EDX and APT measurements are plotted in Fig. 2d. Since EDX is strongly affected by the background signal, the EDX profile was overlapped on the APT profile. A steep transition to ~12 at.% Sn is obtained in just 5-6 nm, followed by a more gradual increase in Sn content toward outer region of the shell with 16 at.%. Considering a lattice mismatch of 2.4 % between $Ge_{0.84}Sn_{0.16}$ and Ge, the absence of extended defects in the nanostructure and the steep compositional profile demonstrates the excellent strain minimization in the system, which behaves as a highly compliant substrate.[28,32,34]

The transition from the conformal core/shell growth to anisotropic two-dimensional fin growth is exemplified in Fig. 3a showing the cross-sectional TEM acquired along the <110> direction of the <111>-oriented NW. The Ge/GeSn core/shell region shows a highly crystalline material (Fig. 3b). Defects are visible in the upper portion of the core/shell in the proximity of the transition region (Fig. 3c), where the streaking in the FFT results from stacking faults (twin boundaries) that are orthogonal to the <111> direction (#1). Moving up along the NW in the transition region, additional stacking faults (twin boundaries) are visible (Fig. 3d). These defects are now, however, orthogonal to the <111> direction (#2). Eventually, the defect density further increases such that it becomes challenging to assign the FFT diffraction spots, and these defects propagate all the way to the top region of the fin (Fig. 3e).

Cross-sectional TEM was performed in the direction perpendicular to the growth axis to evaluate the compositional profile and crystalline quality of the Ge/GeSn core/fin nanostructure. The EDX cross sectional maps of a fin, representative of the general growth process, is shown in Fig. 4a. The fin has a thickness of 30 nm, hence identical to the Ge core, and a width of 80 nm. Interestingly, a Ge-rich stripe is visible in the central region of the GeSn fin, while the remaining portion of the fin reaches a higher, uniform Sn content of around 16 at.%. The measured compositional profile is very similar to that of the Ge/GeSn core/shell region (Fig. 2) despite the absence of GeSn growth along two opposite facets of the Ge core. Performing APT on the fin region is not feasible as the preparation of a tip-shaped specimen is practically impossible, thus the estimate of the fin composition relies mainly on EDX and STEM. We highlight that the higher strain relaxation in the GeSn fins enables a uniform incorporation of Sn through both the longitudinal (Fig. 1k) and lateral (Fig. 4a) directions of the fin. However, unlike the case of the defect-free Ge/GeSn core/shell NWs, plastic strain relaxation occurs in the fins.

To further elucidate the nature of the planar defects along the lateral direction, cross-sectional high-resolution STEM (HRSTEM, Fig. 4b) and HRTEM (4c&d) measurements were carried out. Twin stacking faults extending across the entire width of the fins are visible. Therefore, the strain associated with the 2.4 % lattice mismatch between the Ge core and $Ge_{0.84}Sn_{0.16}$ is relieved by the formation of extended defects in the fins (plastic strain relaxation) and by their inherent small thickness (<30 nm) that promotes elastic strain relaxation at the GeSn growth front. By filtering the HRSTEM images through an inverse-FFT process the GeSn dumbbells become visible with a bond length of 1.6±0.1 Å (Fig. 4e). Despite the difference in atomic numbers between Ge (32) and Sn (50) atoms, given the Sn content is 16 at.% it is not possible to clearly identify individual Sn atoms in the alloy.

Raman measurements were performed on single NWs and nanofins using a 633 nm laser focused into a ~900 nm spot on the sample surface. Figure 5 displays a few representative spectra. Both the NWs and fins exhibit a main peak at around 294 cm$^{-1}$, corresponding to the Ge-Ge mode (F$_2$g mode in diamond lattice) and indicating a compressive strain of -1.5 % in Ge$_{0.84}$Sn$_{0.16}$.[35] The use of 20 nm Ge cores strongly reduces the strain in the shell[26,34] well below the -2.4 % expected for planar Ge$_{0.84}$Sn$_{0.16}$ on Ge. Interestingly, Raman spectra of GeSn fins exhibit also a second peak at 284 cm$^{-1}$, which can be attributed to wurtzite-like regions containing stacking faults, consistent with the ABABAB stacking sequences observed in HAADF images. The peak at 284 cm$^{-1}$ also agrees well with the E$_2$g mode of Lonsdaleite Ge predicted by density functional perturbation theory.[36] The relative intensity of this peak varies between different analyzed GeSn fin structures. Similar behavior was reported in silicon NWs and was attributed to the extent of the regions with stacking faults.[37]

One may invoke multiple factors to explain the mechanisms underlying the transition from conformal GeSn shell to two-dimensional fin growth. First, the diffusion of adatoms from the Ge substrate to the NW sidewall is generally limited to few micrometers and it is strongly reduced with the relatively low growth temperature. The enhanced gas precursor decomposition at the Au droplet (catalytic effect) might compensate for the reduced diffusion length on the sidewall. The interplay between these two factors could result in a change in the growth kinetics in the upper region of the NW that may suppress GeSn growth along two out of six facets of the Ge core. Second, the use of an ultra-thin Ge core of 20 nm in diameter may modify the difference between the surface energy of the (wide) {112} and (narrow) {110} Ge facets. A lower energetic barrier for the Sn incorporation was calculated on the {112} facet compared to the {110} facet,[31] which is indeed observed in both the GeSn shell (Fig. 2) and fins (Fig. 3) growth. The preferential growth of the GeSn fins along the <110> direction cannot be directly correlated with a possible change in surface energy, and it would require more advanced kinetic and atomistic description of the system. Strain is another key element due to the lattice-mismatched nature of the Ge/GeSn core/shell NWs, since it plays a major role in the growth kinetics of GeSn on Ge/Si.[16,38] As the shell grows thicker the strain energy of the system increases, and instabilities along the NW growth axis may lead to the nucleation of defects to reduce strain locally in the shell. The presence of defects may then suppress the GeSn shell growth along two opposite facets and result in the transition into a fin-shaped growth. Despite the complex interplay between surface energy and strain during growth, it is important to highlight the similarities with other material systems, in particular linked to the presence of structural defects in nanoscale structures. The growth of free-standing, defect-free oxide nanobelts was extensively investigated in ZnO.[39] The difference in surface energy between the polar (0001), and non-polar (01-10), (2-1-10) facets was considered to be the main driving force in the nanobelts-nanowires growth transition.[39] Moreover, intentional doping was also pointed out as a growth parameter to engineer the transition from nanowires to nanobelts. Similar results were also obtained in $SnO_2$, $In_2O_3$, and CdO materials,[40] thus highlighting a more general behavior that is associated with energy considerations. A very interesting case is InSb, where the controlled switch from NW to flakes with a thickness of 60 nm was demonstrated to originate from the nucleation of twin plane defects in the bottom region of the NWs.[41] The twinned sections propagate along the entire length of the InSb flakes in a similar fashion to what is observed in the GeSn fins (Figs. 3&5). Indeed, we can identify longitudinal defects (#1 in Fig. 3e) that generate mid-way along the Ge/GeSn nanostructure and that may trigger the nucleation of the GeSn fin. Overall, a plausible interpretation for the shell-fin transition is the local change in growth conditions about 1-2 µm from the bottom of the NWs (diffusion-limited length of Ge,Sn adatoms) that leads to the nucleation of twinning defects in the shell to reduce strain, which then triggers the fin-shaped GeSn growth on four of the {112} facets of the Ge core.

In conclusion, we reported on a spontaneous transition from a conformal shell to quasi-two-dimensional fins during grown of GeSn on ultra-thin Ge nanowires. This transition breaks the 6-fold symmetry typical to core-shell nanowires due to the suppressed growth on two opposing facets. Uniform 16 at.% Sn composition along the lateral <110> and longitudinal <111> directions of the GeSn fins is observed, while preserving the typical higher Sn content on the {112} growing facet compared {110} to the facets of the Ge core. The shell-fin transition is most likely triggered by the local change in growth conditions few µm from the bottom of the NWs, which then induces twinning defects in the GeSn to reduce strain and forces the growth transition into of a fin nanostructure. High-resolution TEM and Raman investigations confirmed that the fins have a polytypic structure consisting of diamond cubic and hexagonal-like regions. We envision harvesting the process to be used in a controlled fabrication of fins arrays, where changing the pitch will enable superior control over the position and length of the fin. This would lead to a nano-engineered material platform that could offer enhanced functionalities compared to top-down under-etched nanomembranes.[42,43]

## ACKNOWLEDGEMENTS

The authors acknowledge support from NSERC Canada, Canada Research Chairs, Canada Foundation for Innovation, Mitacs, PRIMA Québec, and Defense Canada (Innovation for Defense Excellence and Security, IDEaS), the European Union's Horizon Europe research and innovation program under grant agreement No 101070700 (MIRAQLS), and the US Air Force Research Office Grant No. W911NF-22-1-0277.

## AUTHOR INFORMATION

∥ These authors contributed equally to this work.

Corresponding Author:

*E-mail: simone.assali@cea.fr.

Notes: The authors declare no competing financial interest.

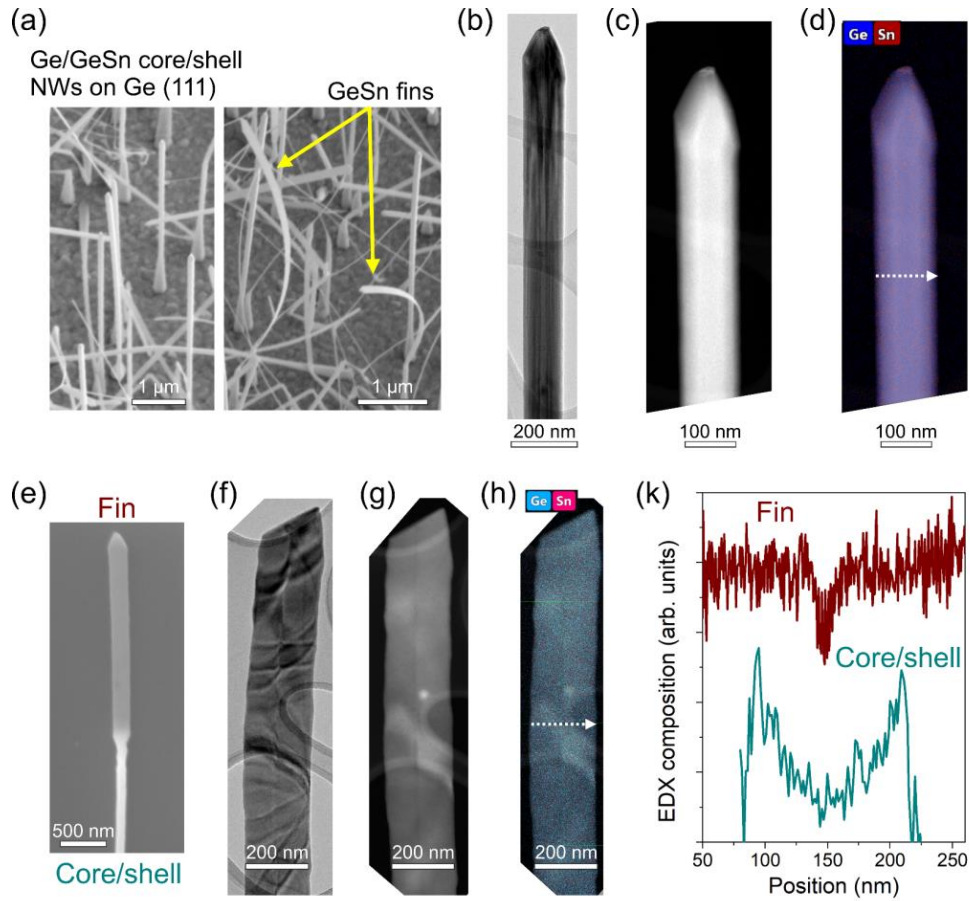

**Figure 1.** (a) SEM image showing the Ge/GeSn core/shell NWs and nanofins grown on a Ge (100) substrate. (b-d) TEM, HRSTEM, and EDX maps of a core/shell NW acquired along the <112> direction. (e) SEM image showing the transition between the shell and the fin for a sample transferred horizontally on a Si substrate. (f-h) TEM, HRSTEM, and EDX maps of a GeSn fin acquired along the <110> direction. (k) EDX line-scan extracted from the direction perpendicular to the NW growth axis for the shell and fin regions.

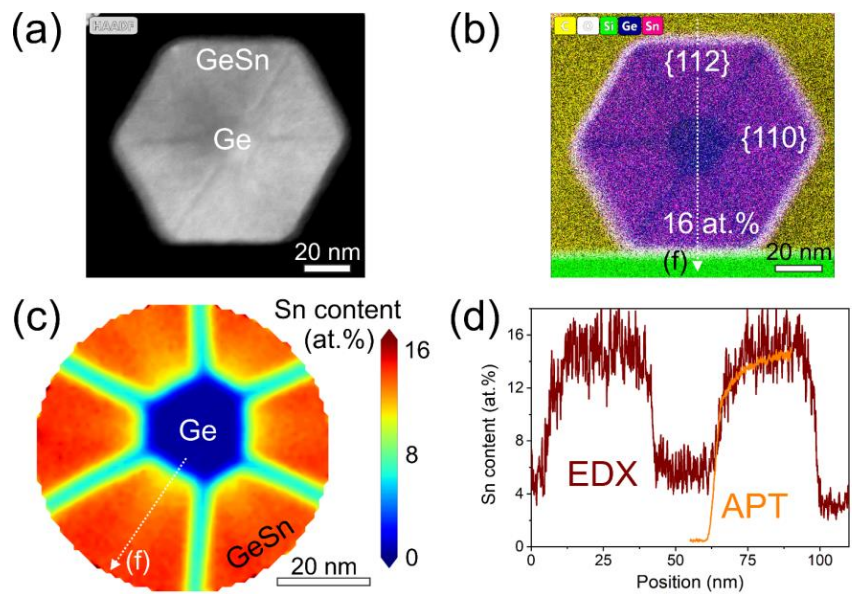

**Figure 2**. (a-b) Cross-sectional STEM (a) and EDX (b) images acquired along the perpendicular axis of a core/shell NW, adapted from Ref.[27]. (c) APT map. (d) Sn compositional profile extracted from the EDX and APT images.

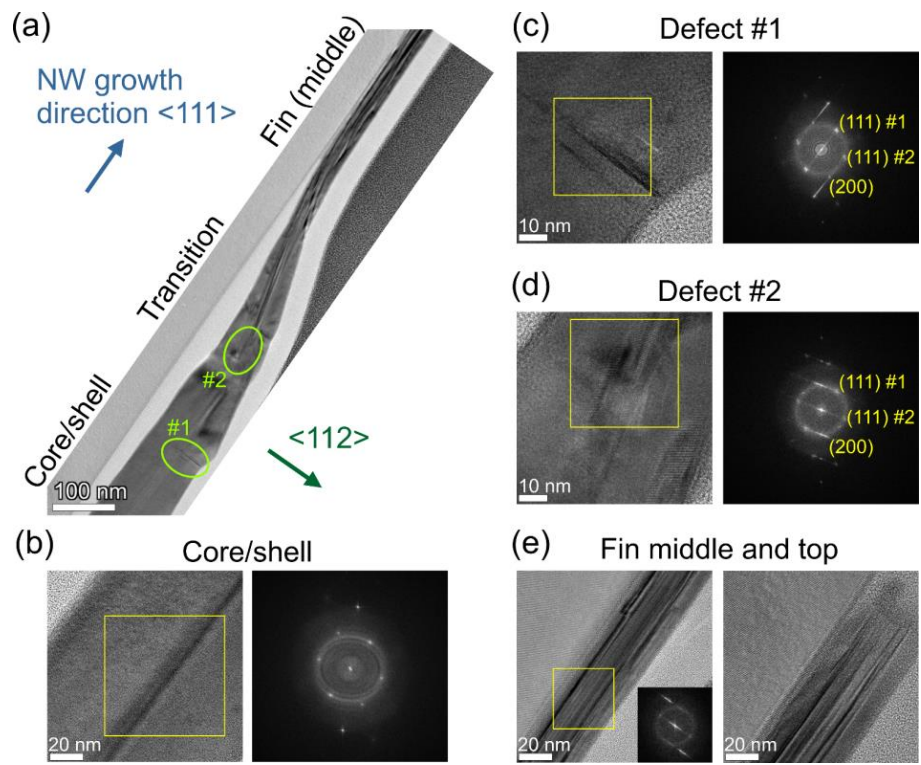

**Figure 3**. (a) Cross-sectional TEM image acquired along the <110> direction of the NW from Figure 1a showing the GeSn shell to fins transition. (b-e) HRTEM images and corresponding FFT acquired along multiple positions of the nanostructure.

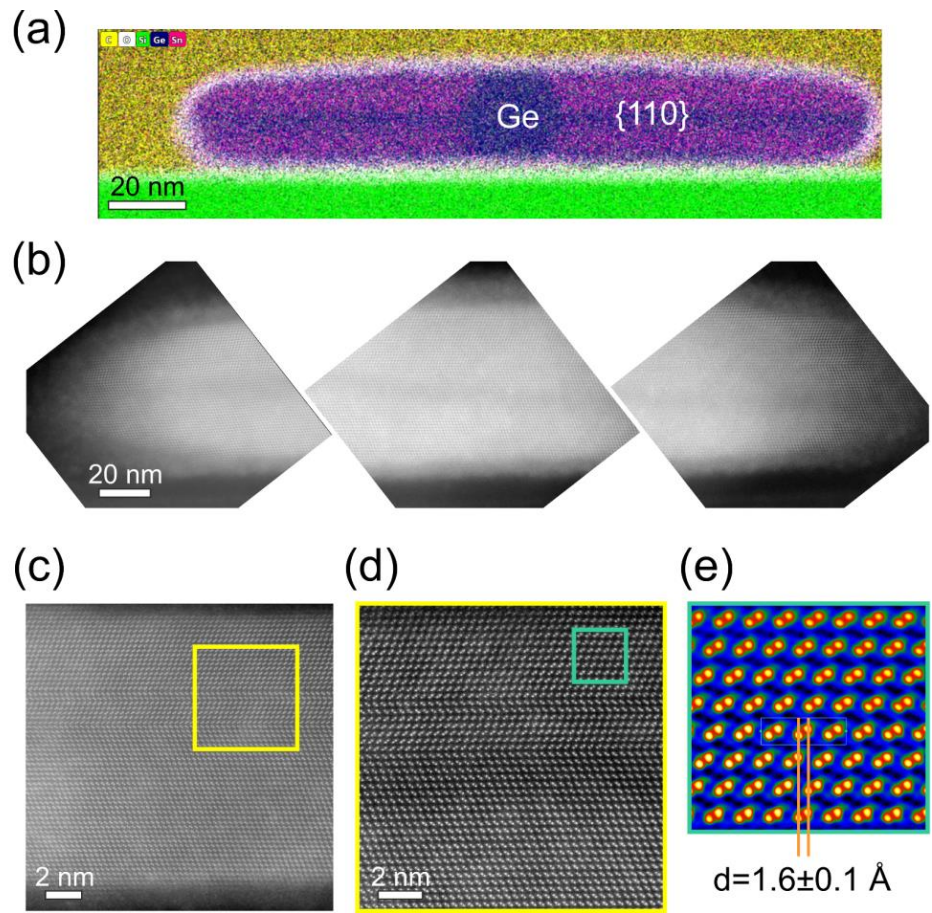

**Figure 4.** (a-b) Cross-sectional EDX (a) and STEM (b) images of a representative GeSn fin. (c-d) HRSTEM images of (b). (e) Filtered image of (d) through inverse-FFT process showing the GeSn dumbbells with a bond length of 1.6±0.1 Å.

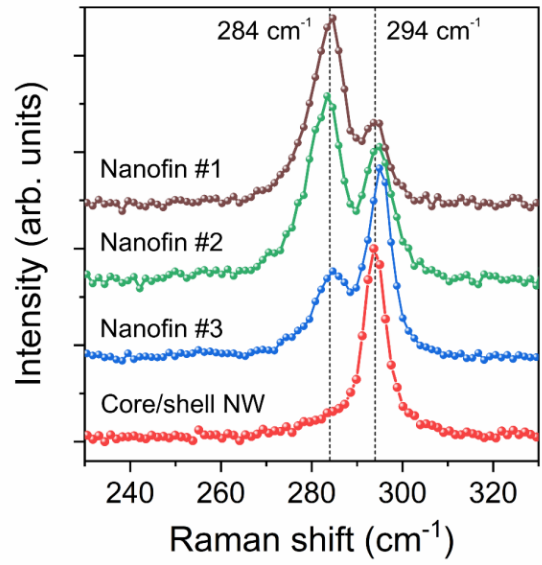

**Figure 5.** Raman spectra acquired on the core/shell and nanofins regions of the sample.